\documentclass[%
 aip,
 apl,%
 amsmath,amssymb,
reprint,%
]{revtex4-2}

\usepackage[x11names]{xcolor}

\usepackage{graphicx}
\usepackage{dcolumn}
\usepackage{bm}
\usepackage{mathrsfs}
\usepackage{float}
\usepackage{hyperref}
\usepackage{url}
\usepackage[normalem]{ulem}

\newif\ifproofread

\usepackage{lineno}

\begin{document}

\proofreadtrue

\title[Characterizing TES Power Noise for Future Single Optical-Phonon and Infrared-Photon Detectors]{Characterizing TES Power Noise for Future Single Optical-Phonon and Infrared-Photon Detectors}

\author{C.W.~Fink}
 \email{cwfink@berkeley.edu.}
\affiliation{Department of Physics, University of California, Berkeley, CA 94720, USA\looseness=-1}

\author{S.L.~Watkins}
\affiliation{Department of Physics, University of California, Berkeley, CA 94720, USA\looseness=-1}

\author{T.~Aramaki}
\affiliation{SLAC National Accelerator Laboratory/Kavli Institute for Particle Astrophysics and Cosmology, Menlo Park, CA 94025, USA\looseness=-1}

\author{P.L.~Brink}
\affiliation{SLAC National Accelerator Laboratory/Kavli Institute for Particle Astrophysics and Cosmology, Menlo Park, CA 94025, USA\looseness=-1}

\author{S.~Ganjam}
\affiliation{Department of Physics, University of California, Berkeley, CA 94720, USA\looseness=-1}

\author{B.A.~Hines}
\affiliation{Department of Physics, University of Colorado Denver, Denver, CO 80217, USA\looseness=-1}

\author{M.E.~Huber} 
\affiliation{Department of Physics, University of Colorado Denver, Denver, CO 80217, USA\looseness=-1}
\affiliation{Department of Electrical Engineering, University of Colorado Denver, Denver, CO 80217, USA\looseness=-1}

\author{N.A.~Kurinsky}
\affiliation{Fermi National Accelerator Laboratory, Batavia, IL 60510, USA\looseness=-1} \affiliation{Kavli Institute for Cosmological Physics, University of Chicago, Chicago, IL 60637, USA\looseness=-1}

\author{R.~Mahapatra}
\affiliation{Department of Physics and Astronomy, and the Mitchell Institute for Fundamental Physics and Astronomy, Texas A\&M University, College Station, TX 77843, USA\looseness=-1}

\author{N.~Mirabolfathi}
\affiliation{Department of Physics and Astronomy, and the Mitchell Institute for Fundamental Physics and Astronomy, Texas A\&M University, College Station, TX 77843, USA\looseness=-1}

\author{W.A.~Page}
\affiliation{Department of Physics, University of California, Berkeley, CA 94720, USA\looseness=-1}

\author{R.~Partridge}
\affiliation{SLAC National Accelerator Laboratory/Kavli Institute for Particle Astrophysics and Cosmology, Menlo Park, CA 94025, USA\looseness=-1}

\author{M. Platt}
\affiliation{Department of Physics and Astronomy, and the Mitchell Institute for Fundamental Physics and Astronomy, Texas A\&M University, College Station, TX 77843, USA\looseness=-1}

\author{M.~Pyle}
\affiliation{Department of Physics, University of California, Berkeley, CA 94720, USA\looseness=-1}

\author{B.~Sadoulet}
\affiliation{Department of Physics, University of California, Berkeley, CA 94720, USA\looseness=-1}

\author{B.~Serfass}
\affiliation{Department of Physics, University of California, Berkeley, CA 94720, USA\looseness=-1}

 \author{S.~Zuber}
\affiliation{Department of Physics, University of California, Berkeley, CA 94720, USA\looseness=-1}

\date{\today}

\begin{abstract}
In this letter, we present the performance of a $100\,\mu\mathrm{m}\times 400\,\mu\mathrm{m} \times 40\,\mathrm{nm}$ W Transition-Edge Sensor (TES) with a critical temperature of $40\,\mathrm{mK}$. This device has a noise equivalent power of $1.5\times 10^{\text{-}18}\, \mathrm{W}/\sqrt{\mathrm{Hz}}$, in a bandwidth of $2.6\,\mathrm{kHz}$, indicating a resolution for Dirac delta energy depositions of $40\pm 5\,\mathrm{meV}$ (rms). The performance demonstrated by this device is a critical step towards developing a $\mathcal{O}(100)\,\mathrm{meV}$ threshold athermal phonon detector for low-mass dark matter searches.
\end{abstract}

\maketitle


As dark matter (DM) direct detection experiments probe lower masses, there is an increasing demand for sensors with excellent energy sensitivity. Several athermal phonon sensitive detector designs have been proposed using superconductors\cite{Hochberg_2016} or novel polar crystals\cite{polar1, polar2,Kurinsky_2019,griffin2019multichannel} as the detection medium. Additionally, experiments that use single infrared (IR) sensitive photonic sensors to read out low band gap scintillators or multi-layer optical haloscopes for both axion and dark photon DM have also been proposed~\cite{haloscope}.

Each of these designs would ultimately require sensitivity to single optical-phonons or IR-photons, corresponding to energy thresholds of $\mathcal{O}(100)~\mathrm{meV}$~\cite{polar1, polar2, Hochberg_2016, haloscope}. Coherent neutrino scattering experiments have made recent progress using DM detector technology and are also interested in cryogenic detectors with very low thresholds~\cite{coh}. Transition-Edge  Sensor (TES) based detector concepts have been successfully applied in DM searches~\cite{CDMSII, single_charge_DM, cresst2019}, as well as IR and optical photon sensors~\cite{NIST_W_TES}. The same concepts can also be used in these new applications, as the necessary energy sensitivities can theoretically be achieved~\cite{Hochberg_2016, polar1}.

The energy resolution of a calorimeter can be estimated with an optimum filter (OF)~\cite{OF, Matt_thesis} from
\begin{align}
    \sigma^2_E = \left[\varepsilon^2 \int_{0}^\infty \frac{d\omega}{2\pi}\frac{4|p(\omega)|^2}{S_P(\omega)}\right]^{\text{-}1},
    \label{eq:OF}
\end{align}
where $S_P(\omega)$ is the total (one-sided) power-referred noise spectrum, $\varepsilon$ is the total phonon collection efficiency, and $p(\omega)$ is power-referred pulse shape defined as ${p(\omega)=1/(1+j\omega \tau_{ph})}$, with $\tau_{ph}$ the athermal phonon collection time of the detector. The resolution for a TES-based calorimeter is minimized when the noise is dominated by the intrinsic thermal fluctuation noise (TFN) between the TES and the bath~\cite{Irwin2005}. This noise can be written as
\begin{align}
   S_{P}(\omega) \approx 4k_BT_c^2G F(T_c, T_B) (1+\omega^2\tau_-^2),
\end{align}
where $k_B$ is the Boltzmann constant, $T_c$ is the superconducting (SC) critical temperature, $T_B$ is the temperature of the bath, $G$ is the dominant thermal conductivity between the TES and the bath, and ${F(T_c, T_B)\approx 1/2}$ is a scale factor accounting for the nonequilibrium nature of the thermal conductance. The effective time constant\footnote{This is also commonly referred to as $\tau_{eff}$ or $\tau_\text{ETF}$} in the strong electrothermal feedback zero inductance limit (also neglecting small effects from the resistance terms and the current sensitivity) can be approximated as ${\tau_{\text{-}} \approx C\sqrt{2n}/(G\alpha)}$, where $\alpha$ is the dimensionless temperature sensitivity, $C$ is the heat capacity, and $n$ is the thermal conduction power law exponent. Under this scenario, the integral in Eq.~(\ref{eq:OF}) becomes
\begin{align}
    \sigma^2_E \approx \frac{1}{\varepsilon^2} 4k_BT_c^2GF(T_c, T_B)(\tau_{ph} + \tau_-).
    \label{eq:eres}
\end{align}

If the energy of an incident particle is absorbed directly by the TES, that is, $\tau_{ph} =0$ and $\varepsilon=1$, then the energy variance in Eq.~(\ref{eq:eres}) becomes
\begin{align}
    \sigma_E^2 \approx k_B T_c^2\frac{C}{\alpha}\sqrt{\frac{n}{2}}\,.
    \label{eq:resTFN}
\end{align}
For a metal in the low-temperature regime, the heat capacity scales with the volume of the TES $(\mathrm{V}_\mathrm{TES})$ and the temperature as ${C(T) \propto \mathrm{V}_{\text{TES}}T}$, suggesting 
\begin{align}
\sigma_E^2 \propto \mathrm{V}_\text{TES} T_c^3. 
\label{eq:tc3}
\end{align}

However, if operated as an athermal phonon sensor, specifically a Quasiparticle-trap-assisted Electrothermal-feedback Transition-edge sensor (QET)~\cite{QET}, the energy sensitivity dependence on $T_c$ becomes even more important. The energy resolution is minimized when athermal phonons bounce in the crystal for times long compared to the characteristic time scale of the TES sensor (i.e. ${\tau_- < \tau_{ph}}$)~\cite{Hochberg_2016, pyle_CRESST, Matt_thesis}, as long as the surface athermal phonon down-conversion rate is negligible~\cite{Knaak}. In this case, the thermal conductance term is not cancelled from Eq.~(\ref{eq:eres}). For low-$T_c$ W films, the thermal conductance is dominated by electron-phonon decoupling, thus scaling as $G\propto \mathrm{V}_\text{TES}T_c^{n-1}$ with $n=5$, as confirmed by measurement described later in this letter. This implies that the baseline energy variance of the detector will scale with critical temperature as $\sigma_E^2\propto T_c^6$, suggesting that a low-$T_c$ device is ideal for single optical-phonon sensitivity.

\begin{figure}
\includegraphics[width=\linewidth]{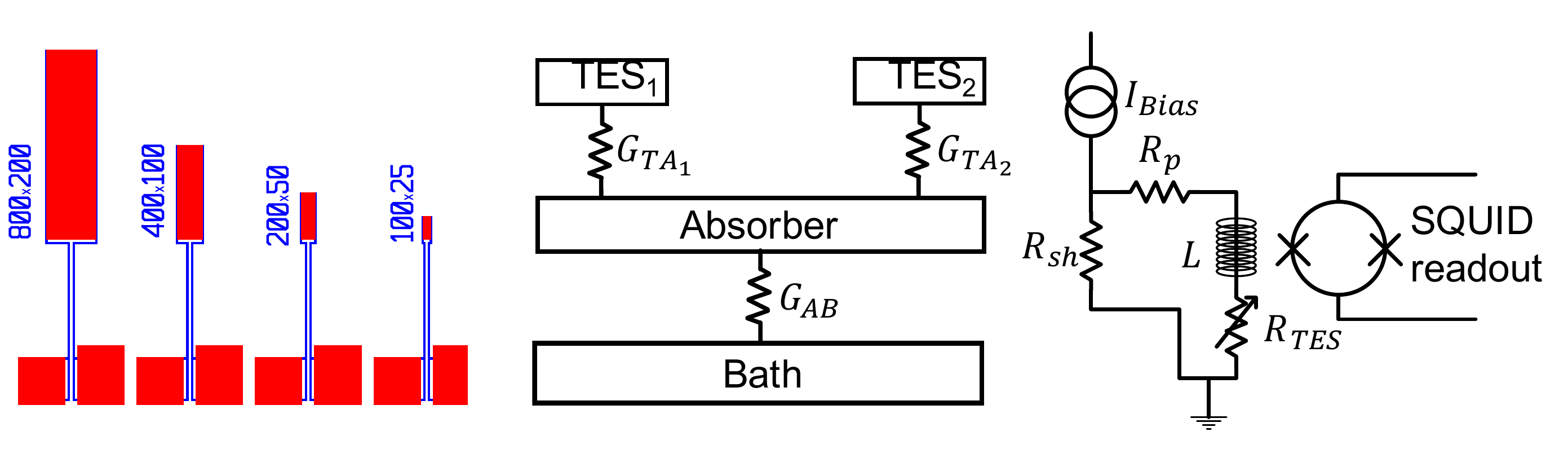}
\caption{Left: TES mask design. The W is shown in red, while the blue represents Al bias rails. The Al connects to the left and right sides of the TES. Middle: Thermal model for experimental setup. For simplicity, only two TESs are shown in the model. Right: Electrical circuit. $R_{sh}$ is a shunt resistor which turns the current source ($I_\text{Bias}$) into a voltage bias. Any parasitic resistance on the shunt side of the bias circuit is absorbed into the value used for $R_{sh}$ in this analysis. $R_p$ is the parasitic resistance on the TES side of the bias circuit. $L$ is the inductance in the TES line. $R_\text{TES}$ is the TES resistance, which takes on a value of $R_0$ when in transition and takes on a value of $R_N$ when its temperature is above $T_c$.}
\label{fig:setup}
\end{figure}

A set of 4 W TESs was fabricated on a $525 \,\mu\mathrm{m}$ thick $1\,\mathrm{cm}\times 1\,\mathrm{cm}$ Si substrate (``chip''). The smallest of the TESs was $25\,\mu \mathrm{m} \times 100\,\mu \mathrm{m} \times 40\,\mathrm{nm}$. Each subsequent TES increased in area by a factor of four, keeping an aspect ratio of 1:4 (width\,:\,length), which implies all the TESs have the same normal resistance ($R_N$). The TES mask design can be seen in left panel Fig. \ref{fig:setup}. Two sets of these chips were made, one with TESs of ${T_c=40\,\mathrm{mK}}$ and the other with TESs of $T_c=68\,\mathrm{mK}$. This letter focuses on the measurement and characterization of the low-$T_c$ $100\,\mu \mathrm{m} \times 400\,\mu \mathrm{m} \times 40\,\mathrm{nm}$ TES (hereby referred to as simply ``the TES''), but will also present characterization data from these other devices to elucidate scalings with $T_c$ and volume. The utility of such devices for applications of photon detectors and athermal phonon sensors will also be discussed.

The voltage-biased TES was studied at the SLAC National Accelerator Laboratory in a dilution refrigerator at a bath temperature of $15\,\mathrm{mK}$. The Si chip was mounted to a copper plate with GE varnish. The current through the TES was measured with a custom DC Superconducting Quantum Interference Device (SQUID) array system with a noise floor of $\sim\!4~\mathrm{pA}/\sqrt{\mathrm{Hz}}$, fabricated for the SuperCDMS experiment, with a measured lower bound on the bandwidth of greater than $250\,\mathrm{kHz}$. The SQUID array was read out by an amplifier similar to the one in Ref.~\onlinecite{revC}.

Multiple measures were put in place to mitigate electromagnetic interference (EMI). Pi-filters with a cutoff frequency of $10\,\mathrm{MHz}$ were placed on all input and output lines to the refrigerator. Ferrite cable-chokes were placed around the signal readout cabling at $300\,\mathrm{K}$, and the 4K and 1K cans were filled with broadband microwave-absorptive foam to suppress radio frequency (RF) radiation onto TESs. The outer vacuum chamber of the dilution refrigerator was surrounded by a high-permeability metal shield to suppress magnetic fields. These measures were the result of a systematic search of the system's susceptibility to environmental noise, and they lowered the measured electrical noise by roughly an order of magnitude. Despite these efforts, an unknown parasitic noise source remained, which inhibited the smallest two low-$T_c$ TESs from going through their SC transition.

To characterize the TES, $IV$ sweeps were taken at various bath temperatures by measuring TES quiescent current ($I_0$) as a function of bias current ($I_\text{Bias}$)\footnote{We use the term ``IV'' even though we are applying a bias current, as the voltage and current are related by the shunt resistor: $V_\text{Bias} = I_\text{Bias}R_{sh}$}, with complex admittance data taken at each point in the $IV$ curve~\cite{Matt_thesis, Noah_thesis}. Data were also taken simultaneously with the largest low-$T_c$ TES (TES2) on the same Si chip, biased at an operating resistance ($R_0$) of approximately  $40\%\,R_N$, in order to attempt to quantify the amount of remaining excess noise that coupled coherently to both TES channels. From the $IV$ sweep at each temperature, both the DC offset from the SQUID and any systematic offset in $I_\text{Bias}$ were corrected for using the normal and SC regions of the data. After this correction, the parasitic resistance in the TES circuit ($R_p$), the normal state resistance, the TES resistance in transition, and the quiescent bias power ($P_0$) were calculated (see the right panel of Fig.~\ref{fig:setup} for circuit diagram).

Since the Si chip contained multiple TESs, the thermal conductance between the chip and the bath ($G_{AB}$) was measured by using one as a heater and one as a thermometer. Knowledge of $G_{AB}$ allowed us to infer the temperature of the Si chip ($T_A$) from a measurement of $T_B$. See the middle panel of Fig.~\ref{fig:setup} for a thermal diagram of the setup. Measuring $P_0$ as a function of temperature from the $IV$ sweeps, the thermal conductance between the TES and the Si substrate ($G_{TA}$), $T_c$, and $n$ were fit  to a power law~\cite{Electron_Phonon_Coupling}, confirming our $n=5$ assumption. We measured that $G_{AB}$ was roughly 3 orders of magnitude larger than $G_{TA}$, meaning that $T_A$ was effectively equal to $T_B$, and the system could be modeled as a single thermal conductance between the TES and the bath. The characteristics of the TES system from the $IV$ data are shown in Table~\ref{tab:rp}.

\begin{table*}
    \centering
     \caption{Various calculated parameters of the TES. $R_{\square}$ or ``R-square" is the sheet resistance of the W film.}
    \begin{tabular}{|c|c|c|c|c|c|c|c|c|c|c|}
    \hline
    $R_{sh}\,[\mathrm{m}\Omega]$ & 
    $R_p\,[\mathrm{m}\Omega$] & $R_N\,[\mathrm{m}\Omega$] & $R_{\square}\, [\Omega]$ & $P_0\, [\mathrm{fW}]$  & $G_{AB}\, [\mathrm{nJ}/\mathrm{K}]$ & $G_{TA}\, [\mathrm{pJ}/\mathrm{K}]$  &  $T_c\, [\mathrm{mK}]$ & $T_B\, [\mathrm{mK}]$ & $T_\ell\,[\mathrm{mK}]$ & n\\
    
    \hline
    $5.0\pm0.5$ & 
    $5.8\pm 0.6$ & $640 \pm 65$ & $2.56 \pm 0.26$ & $31 \pm 2$ & $1.6\pm0.1$ & $4.0 \pm 0.4$  &
     $40\pm1$ & $15\pm 1$ & $37\pm 2 $ & 5\\
   \hline
   \end{tabular}
 
   \label{tab:rp}
\end{table*}


For each point in transition, a maximum likelihood fit of the complex admittance was done, using the standard small-signal current response of a TES~\cite{Irwin2005}:
\begin{align}
\begin{split}
    Z(\omega) &\equiv R_{sh} + R_p +j\omega L + Z_{\mathrm{TES}}(\omega), \\
    Z_{\mathrm{TES}}(\omega) &\equiv R_0(1+\beta)+\frac{R_0\mathscr{L}}{1-\mathscr{L}}\frac{2+\beta}{1+j\omega \frac{\tau}{1-\mathscr{L}}}. 
    \label{eq:ztes}
    \end{split}
\end{align}

\begin{figure}
  \includegraphics[width=\linewidth]{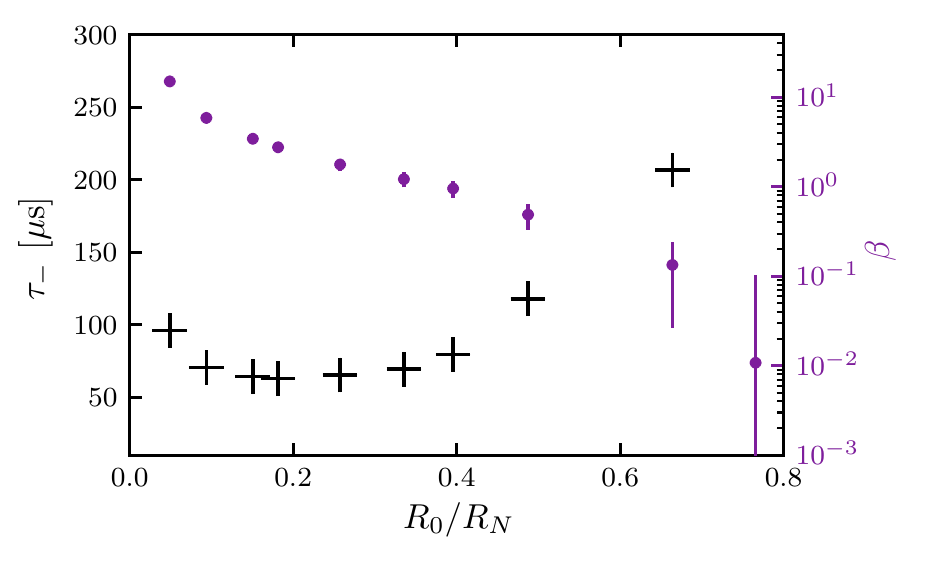}
  \caption{Fitted values for $\beta$ (purple dots) and effective electrothermal TES response time $\tau_-$ (black crosses) as a function of TES resistance.}
  \label{fig:beta_tau}
\end{figure}

\begin{figure}
\includegraphics[width=\linewidth]{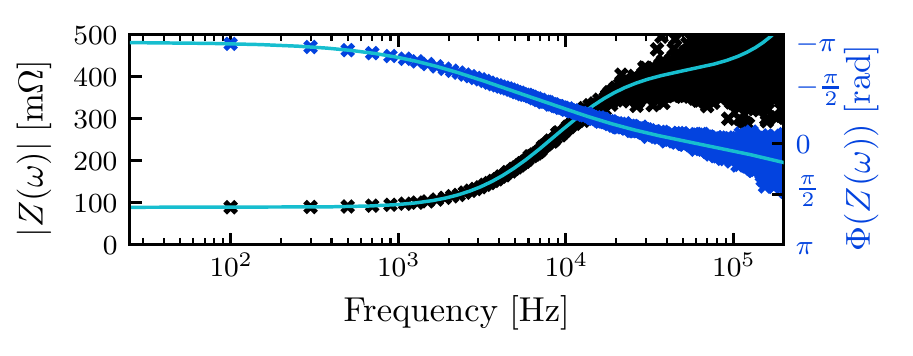}
\caption{A typical complex impedance curve for the TES in transition for $R_0\approx 15\% R_N$. The measured magnitude and phase of the complex impedance are shown in black and blue, respectively. In cyan, the complex impedance derived from the maximum likelihood fitting routine is shown.}
\label{fig:didv}
\end{figure}

In this fit, $L$, $R_0$, $R_p$, $R_{sh}$\footnote{$R_{sh}$, is a free parameter in the fit because we do not have a good measurement of it at cryogenic temperatures.}, $\beta$, $\tau$, and $\mathscr{L}$ are all free parameters. $L$ is the inductance in the TES bias circuit, $\beta$ is the dimensionless current sensitivity, $\tau$ is the natural thermal time constant, and $\mathscr{L}$ is the loop gain. We include the estimates from the $IV$ data of $R_0,\ R_p,\ \text{and} \ R_{sh}$ as priors in the fit. Additionally, we include a prior on $L$, measured from SC complex admittance data. The TES response times can also be measured from the complex admittance data, defined as the rise and fall times of the TES response from a delta function impulse ($\tau_+$ and $\tau_-$, respectively)~\cite{Irwin2005}. Best fit values of $\beta$ and $\tau_-$ are shown in Fig.~\ref{fig:beta_tau}, while a typical complex impedance curve can be seen in Fig.~\ref{fig:didv}.

\begin{figure}
  \includegraphics[width=\linewidth]{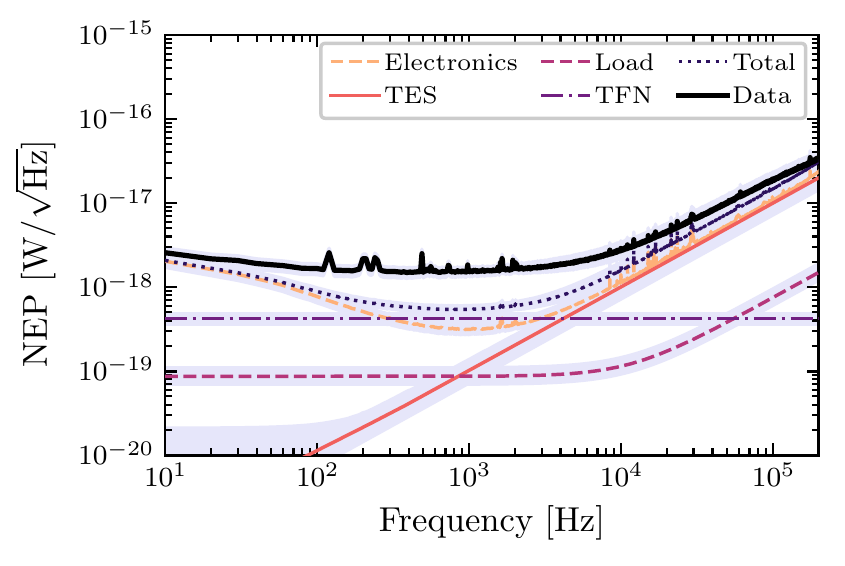}
  \caption{Modeled noise components: TES Johnson noise (orange solid), load resistor Johnson noise (red dashed), electronics noise (yellow dashed), thermal fluctuation noise (purple alternating dashes and dots), and total modeled noise (purple dots) compared with the derived NEP (black solid). The noise model and NEP are shown for $R_0 \approx 15\% R_N$. The shaded regions represent the $95\%$ confidence intervals.}
  \label{fig:transition_noise}
\end{figure}

The normal-state noise was used to estimate the SQUID and amplifier noise, once the Johnson noise component of the TES at $R_N$ was subtracted out. The effective load resistance temperature\footnote{The load resistance is $R_\ell = R_{sh} + R_p$. When the TES is SC, the noise spectrum is dominated by the Johnson noise of the $R_\ell$, $S_{I_\ell} = 4k_B T_\ell R_\ell \left|1/(R_\ell + j\omega L)\right|^2$. With $R_\ell$ and $L$ known, the measured noise can be used to estimate $T_\ell$.} was estimated from the SC noise spectrum, resulting in $T_\ell \approx 37\,\mathrm{mK}$, which was used to estimate the Johnson noise from $R_{sh}$ and $R_p$. The TFN and TES Johnson noise components of the system were calculated as defined in the standard small-signal noise model~\cite{Irwin2005}, using the complex admittance fit parameters. The measured power spectral density (PSD), referenced to TES current, of the device in transition was converted into the noise equivalent power (NEP) with the power-to-current transfer function~\cite{Irwin2005}
\begin{align}
    \frac{\partial I}{\partial P}(\omega) = \left[I_0 \left(1-\frac{1}{\mathscr{L}}\right)\left(1+j\omega\frac{\tau}{1-\mathscr{L}}\right)Z(\omega)\right]^{\text{-}1},
    \label{eq:didp}
\end{align}
where $Z(\omega)$ is defined in Eq.~(\ref{eq:ztes}). A comparison of the noise model to the derived NEP for a typical operating point in transition is shown in Fig.~\ref{fig:transition_noise}.

\begin{figure}
  \includegraphics[width=\linewidth]{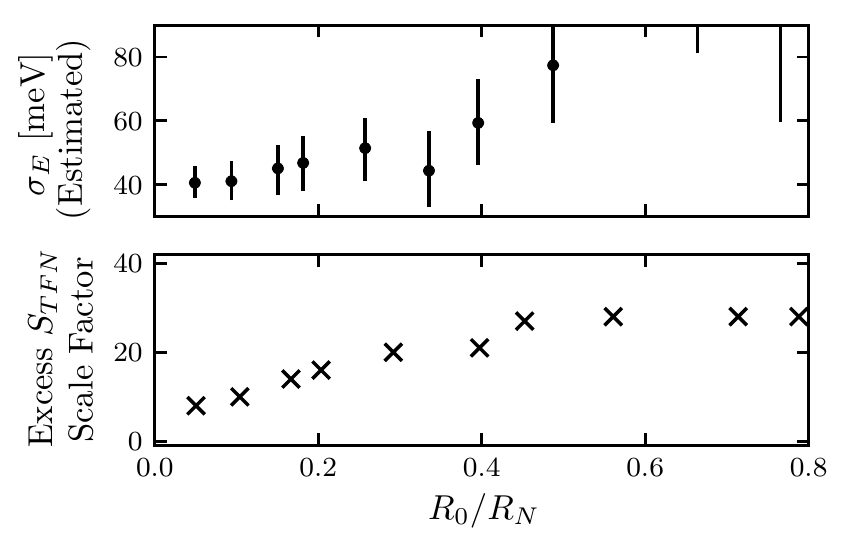}
  \caption{Upper: Estimated energy resolution (from data) throughout the SC transition. Lower: Scale factor needed to increase $S_{TFN}$ to make the noise model match the measured PSD.}
  \label{fig:eres_scale_noise}
\end{figure}

From the derived NEP, the energy resolution of a Dirac delta impulse of energy directly into the TES was estimated using Eq.~(\ref{eq:OF}), with $\varepsilon=1$ and $\tau_{ph}=0$. It can be seen in the upper panel of Fig. \ref{fig:eres_scale_noise} that when the TES is operated at less than $\sim\! 15\% \ R_N$, the estimated resolution of the collected energy is $\sigma_E = 40 \pm 5\,\mathrm{meV}$. At this point in the transition, the sensor has an NEP of $1.5\times 10^{\text{-}18}\,\mathrm{W}/\sqrt{\mathrm{Hz}}$ in a bandwidth of $2.6\,\mathrm{kHz}$. This resolution represents the lower limit of the performance of this sensor given the measured noise, operated as either a photon or athermal phonon sensor. In the case of the athermal phonon sensor, there would be an additional efficiency factor based on the design of the detector.

It is evident from Fig. \ref{fig:transition_noise} that the NEP is elevated from the theoretical expectation across the full frequency spectrum. We split the excess noise into two categories. Noise that scales with the complex admittance and is present when the TES is biased in its normal or SC state, we call ``voltage-coupled", e.g. inductively coupled EMF. Noise that is only seen when the TES is in transition is referred to as ``power-coupled", e.g. IR photons radiating onto device. The excess voltage-coupled noise ($S_{SC^*}$) can be modeled by scaling the SC power spectral density (PSD) by the complex admittance transfer function when the TES is in transition via Eq.~(\ref{eq:sc_scaling}). This modeled noise can then be subtracted from the transition state PSD in quadrature. 
\begin{align}
    S_{SC^*}(\omega) = S_{SC}(\omega) \frac{\left|\left[Z(\omega)\right]_{R_0}\right|^2}{\left|\left[Z(\omega)\right]_{R_0\to 0}\right|^2}
    \label{eq:sc_scaling}
\end{align}

\begin{figure}
  \includegraphics[width=\linewidth]{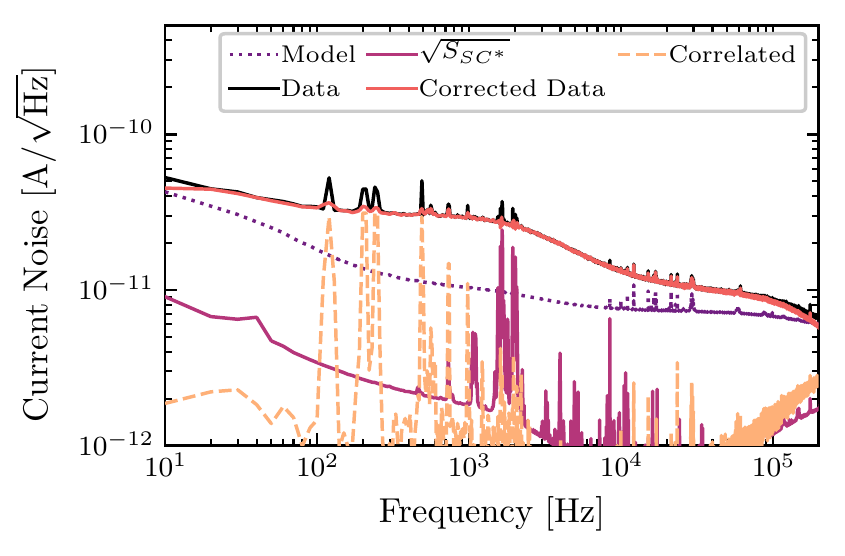}
  \caption{Measured noise (black solid), modeled voltage-coupled noise (purple solid), correlated noise (yellow dashed), measured noise with voltage-coupled and correlated components subtracted (orange solid), and theoretical noise model (purple dots) shown for $R_0\approx 15\% R_N$. The environmental noise model explains the peaks in the measured spectrum, but there is still a discrepancy between the environmental-noise-corrected data and the noise model.}
  \label{fig:corrected_noise}
\end{figure}

We expect power-coupled noise from an environmental origin to couple coherently to each TES on the same Si chip, though we have seen evidence of power-coupled noise generated by the Ethernet chip on our warm electronics to have significantly different couplings to different electronics channels. Because we acquired data simultaneously on TES2, we can determine the correlated and uncorrelated components of the noise by using the cross spectral density (CSD)~\cite{noise_decor, Noah_thesis}. The scaled SC noise PSD and correlated part of the CSD are plotted with the measured PSD in Fig. \ref{fig:corrected_noise} for $R_0\approx 15\% R_N$. The two noise sources can explain the peaks in the noise spectrum, but cannot explain the overall elevated noise level.

To investigate the hypothesis of the excess noise being explained by IR photons radiating onto the TES structure, we modeled this system by multiplying the TFN by a scalar in order to make the total noise model match the NEP. This scale factor is shown in the lower panel of Fig. \ref{fig:eres_scale_noise}. The fact that this scale factor is monotonically increasing with $R_0$ implies that this mechanism is not a dominant source of excess noise, as it should be independent of the TES operational bias point. 

We ruled out the possibility of the excess noise being due to multiple thermal poles~\cite{multipole_noise, 2pole_thermal_model}, as none of these models were able to explain the observed noise spectra. This is also evident by noting the lack of additional poles in the complex impedance in Fig. \ref{fig:didv}.
 
The fact that the two smallest low-$T_c$ TESs (the most sensitive to parasitic power noise) were not able to go through their SC transition, suggests that a nonnegligible amount of the excess noise is environmental in origin. However, given the previous discussion, this leaves open the possibility that some of this excess noise is intrinsic to the TESs.

We compare the estimated energy resolution of the TES to the high-$T_c$ TESs, using the same analysis techniques, in Table~\ref{tab:high_tc}. The high-$T_c$ TESs also observed a similar amount of excess noise. Despite the elevated noise seen on both sets of TESs, the resolution scaling with volume and $T_c$ from Eq.~(\ref{eq:tc3}) still approximately holds. We note that we do not compare the energy resolutions using the expected scaling relation for athermal phonon sensors because of its dependence on both substrate material and QET geometry.

\begin{table}
    \centering
    \caption{Energy resolution estimates for $68\,\mathrm{mK}$ $T_c$ TESs compared to the $40\,\mathrm{mK}$ $T_c$ TES described in this work.}
        \begin{tabular}{|c|c|c|c|}
    \hline
          $T_c$ & TES Dimensions  &  $\sigma_E$ & $\sigma_E$\footnotemark[1]\\
              $[\mathrm{mK}]$ & $[\mu\mathrm{m}\times\mu\mathrm{m}\times\mathrm{nm}]$  & $[\mathrm{meV}]$  & $[\mathrm{meV}]$\\
              & & Estimated & Predicted\\
              & & & using Eq.~(\ref{eq:tc3})\\            \hline 
              \hline
            $40$ & $100\times 400\times 40$ & $40\pm 5$ & N/A\\
            
            $68$ & $50\times 200\times 40$ & $44\pm 5$ & $44\pm 5$\\
            
            $68$ & $100\times 400\times 40$ & $104\pm 10$ & $89\pm11$\\

    \hline
    \end{tabular}
     \footnotetext[1]{The resolution expected from a hypothetical device (with the same physical properties) by scaling the resolution of the low-$T_c$ TES ($\sigma_1$) using Eq.~(\ref{eq:tc3}), i.e. $\sigma_x = \sigma_1\sqrt{\mathrm{V}_x T_{c_x}^3/\mathrm{V}_1T_{c_1}^3}$}
    
    \label{tab:high_tc}
\end{table}

With an estimated energy resolution of ${40\pm 5\,\mathrm{meV}}$~(rms), this device has comparable energy sensitivity to world leading optical and near-IR TESs, but with a volume that is much larger, due to its low-$T_{c}$ (see Table \ref{tab:tes_comp}). It has immediate use as a photon detector in optical haloscope applications \cite{haloscope}. Furthermore, its large volume suggests that significant improvements in sensitivity can be made in short order; a $20\,\mu\mathrm{m}\times 20\,\mu\mathrm{m}\times 40\,\mathrm{nm}$ TES made from the same W film would be expected to have $4~\mathrm{meV}$~(rms) sensitivity, provided that we can reduce observed excess noise and the volume scaling in Eq.~(\ref{eq:tc3}) continues to hold.

\begin{table}
    \centering
    \caption{Performance of state-of-the-art TES single photon calorimeters/bolometers.}
    \begin{tabular}{|c|c|c|c|c|c|}
    \hline
         TES & $T_c$ & $V_\mathrm{TES}$ & $\sigma_E$  &  $\frac{\sigma_E}{\sqrt{V_{\mathrm{TES}}}}$ & Method \\
             & $[\mathrm{mK}]$ &  $[\mu\mathrm{m}^3]$ & $[\mathrm{meV}]$ & $\left[\frac{\mathrm{meV}}{\mu \mathrm{m}^{3/2}}\right]$ & \\
            \hline \hline
            
            W\cite{doi:10.1063/1.1596723} & 125  & 21.88 & 120 & 25.7 & measured\\
            Ti\cite{918addfc209841dca7f245dfe9288a30} & 50  & 0.13 & 47 & 128.2 & measured\\

            MoCu\cite{Goldie} & 110.6  &2000 & 295.4 & 6.6 & estimated\footnotemark[2]\\
            TiAu\cite{INRIM} & 106 & 90 & 48 & 16 & measured \\
            TiAu\cite{spica} & 90 & 202.5 & $\sim\!23$ & $1.6$ & estimated\footnotemark[2] \\ 
            \hline
            W (this) & 40 & 1600& 40 & 1& estimated\\
    \hline
    \end{tabular}
    \footnotetext[2]{The energy resolution is estimated with Eq.~(\ref{eq:OF}) from the given NEP and sensor bandwidth.}
    \label{tab:tes_comp}
\end{table}

For athermal phonon sensor applications~\cite{Hochberg_2016,polar1, polar2,Kurinsky_2019,griffin2019multichannel}, the expected resolution is also impacted by the athermal phonon collection efficiency, which is typically $>\,20\%$ in modern designs~\cite{hvev2}. Thus, small-volume crystal detectors ($\sim\!1\,\mathrm{cm}^3$) should be able to achieve sub-eV triggered energy  thresholds. Though such devices could not achieve the ultimate goal of single optical-phonon sensitivity, they could achieve the intermediate goal of sensitivity to single ionization excitations in semiconductors without E-field amplification mechanisms \cite{single_charge_DM, CCD}, which have historically correlated with spurious dark counts. A decrease in TES volume and $T_{c}$, along with concomitant improvements in environmental noise mitigation and the use of crystals with very low athermal phonon surface down-conversion, would additionally be necessary to achieve optical phonon sensitivity. As we expect the energy variance to go as $T_c^6$ in this application, the benefit of lower $T_c$ should be significant.


This work was supported by the U.S. Department of Energy under contract numbers KA-2401032, DE-SC0018981, DE-SC0017859, and DE-AC02-76SF00515, the National Science Foundation under grant numbers PHY-1415388 and PHY-1809769, and Michael M. Garland. The main findings of this letter can be replicated from the presented data, but the full data that support the findings of this study are available from the corresponding author upon reasonable request.

\bibliographystyle{aipnum4-1}
\bibliography{main}

\end{document}